# Improving the Josephson energy in High-T$_c$ superconducting junctions for ultra-fast electronics

Navarro Henry[1*,†], Sirena Martin[1,2], Haberkorn Nestor[1,2].

[1]Instituto Balseiro, Universidad Nacional de Cuyo and Comisión Nacional de Energía Atómica, Av. Bustillo 9500, 8400 San Carlos de Bariloche, Argentina.

[2]Comisión Nacional de Energía Atómica and Consejo Nacional de Investigaciones Científicas y Técnicas, Centro Atómico Bariloche, Av. Bustillo 9500, 8400 San Carlos de Bariloche, Argentina.

We report the electrical transport in vertical Josephson tunnel junctions (area 400 µm2) using GdBa$_2$Cu$_3$O$_{7-\delta}$ electrodes and SrTiO$_3$ as an insulating barrier (with thicknesses between 1 nm and 4 nm). The results show Josephson coupling for junctions with SrTiO$_3$ barriers of 1 nm and 2 nm. The latter displays a Josephson of 8.9 mV at 12 K. This value is larger than the usually observed in planar arrays of junctions. Our results are promising for the development of superconducting electronic devices in the terahertz regime.

The development of tunnel Josephson junctions (JJ) is of technological relevance for many applications going from superconducting quantum interference devices (SQUIDS), to radiation detectors and rapid single flux quantum (RSFQ) digital electronic circuits[1,2,3,4,5]. Most of them require close-packed arrays of junctions, whose characteristic critical current $I_c$ and normal state resistance $R_n$ have to be uniform over large areas and on large scales from a few to tens of thousands of junctions. The use of high-$T_c$ superconductors (HTS) in JJ, due to the large superconducting gap and operating temperature range[6], offers numerous advantages over other superconducting materials such as Nb[7], nitrides[8] and $MgB_2$[9,10]. The large superconducting gap corresponds to a higher Josephson energy, increasing the devices frequency operation rates, relaxing the cooling requirements and reducing significantly the influence of the thermal noise. Most research in JJ based on HTS such as $YBa_2Cu_3O_{7-\delta}$ is oriented to planar arrays developed using ion irradiation [11,12], edge ramp[13,14] and bicrystals[15,16]. The design and fabrication of vertical JJs allows an increased functionality by using barriers with different electronic and magnetic properties. Nevertheless, few successful reports have been published to this date. The main obstacle for the design of vertical tunnel junctions (TJ) using HTS is related to the 3D growth mechanism usually observed in thin films.

In this letter, we report the electrical properties of superconductor/ insulator /superconductor (SIS) vertical TJ's using $GdBa_2CuO_3O_7$ (GBCO) as superconducting electrodes and $SrTiO_3$ (STO) as the insulating barrier. The crystalline structure of GBCO/STO/GBCO trilayers was previously discussed in ref.[17]. The heterostructures display single phase with (00*l*) orientation. The STO surface displays steps which are mainly originated by the roughness of the bottom GBCO electrode (typical 1unit cell ~1.2 nm)[18]. The electronic homogeneity of the STO barrier as a function of the thickness was previously analyzed in GBCO/STO bilayers by conducting atomic force microscopy at room temperature[19]. Conductivity maps indicate higher values at the borders of topological defects than in the terraces. The superconducting critical temperature ($T_s$) of 16 nm thick GBCO films is reduced ($T_s$ ~77 K) in comparison with bulk specimens ($T_s$ ~90 K). Moreover, for GBCO/insulator bilayers the $T_s$ value decreases from ~77 K to ~40 K when the barrier thickness increases from 1 nm to 4 nm[20]. The reduction in $T_s$ could be related mainly with the stress induced by the substrate and the barrier, probably modifying the oxygen content[21]. In our research, the GBCO thickness has been selected by considering a trade-off between $T_s$ and the presence of smooth surfaces[16,20].

The epitaxial GBCO/STO/GBCO trilayers were grown on STO using the conditions previously described in refs.18 and 19. The substrate temperature was kept at 730°C in an Ar (90%)/$O_2$(10%) mixture at a pressure of 400 mTorr. After deposition, the temperature was decreased to 500°C, and the $O_2$ pressure was increased to 100 Torr. Finally, the samples were cooled down to room temperature at a rate of 1.5° C/min. A 2 nm thick STO buffer layer was introduced to reduce the nucleation of 3D topological defects in the GBCO bottom electrode. The tunnel junctions design used 16 nm thick GBCO electrodes and a STO barrier with thickness ($d_{STO}$) of 1 nm, 2 nm, 3 nm, and 4 nm. Wherever used, the notations [G-$d_{STO}$-G] indicate a GBCO bottom and top electrodes and a STO barrier with thickness $d$ (nm). The results presented in this work correspond to TJs with an area of 400 μm$^2$. The fabrication process includes successive steps of optical lithography and Argon ion milling[22]. To avoid electrical shorts while contacting the samples, the TJs were covered with a 100 nm thick $SiO_2$ layer.

Characteristic current-voltage (IV) curves were measured by using the standard four-point geometry in closed cycle refrigerator with a base temperature of 12 K. The current was applied using a Keithley source. The magnetic field $H$ parallel to the barrier surface was applied with a copper coil.

Figure 1 shows IV curves taken in the four studied samples for temperatures ranging from 12 K to 100 K. The slope of the IV curves (emphasized by the dotted line) corresponds to the normal-state junction resistance $R_n$. Josephson coupling is observed for STO barriers with thicknesses of 1 nm and 2 nm. The IV curves for [G-3-G] show features related to superconductivity below 35 K, however as we will discuss below, the effect does not correspond to Josephson coupling. The IV curves in [G-4-G] show features usually observed in metal-insulator-metal junctions[17]. The $R_n$ values are 1.6 Ω [G-1-G], 99 Ω [G-2-G] and 550 Ω [G-3-G], indicating a significant increment in the resistance of the junctions when the STO thickness is increased. The gradual reduction of the $T_s$ with the increment of the barrier thickness, previously reported in bilayers[20], is also manifested in the Josephson coupling temperature ($T_J$), ranging from ~77 K for [G-1-G] to ~35 K for [G-3-G]. It is important to note that the IV curves for [G-1-G] show hysteresis, which is usually associated with an underdamped behavior, while the IV curves for [G-2-G] do not show hysteresis, corresponding to the overdamped behavior. For the latte, due to a higher Josephson energy, a reduced influence of thermal fluctuations leads to smaller variations in the switching behavior.

In order to verify the presence of Josephson coupling, we investigated the effect of the external magnetic field $H$ perpendicular to the Josephson current. For that, we performed IV curves applying different magnetic fields. The expected field dependence of the $I_c$ in the Fraunhofer patterns for rectangular Josephson junctions is given by:

$$I(H) = I_0 \left| \frac{\sin(\frac{\pi\Phi}{\Phi_0})}{\frac{\pi\Phi}{\Phi_0}} \right|, \text{[eq. 1]}$$

here $I_0 = J_1WL$ ($J_1$ the total current density, $W$ and $L$ the lengths of the junction), $\phi_0 = 2.07 \cdot 10^{-7}$ G·cm$^{-2}$, $\phi = H \cdot L \cdot d$ (with $d$ being the effective junction thickness). Figures 2 *a* and *b* show the $I_c$ versus $H$ dependences for [G-1-G] and [G-2-G]. The measurements for [G-3-G] are not included because the $I_c$ in this sample was not affected by the field. There are two typical characteristics to be considered in the Fraunhofer patterns observed for these two samples. The first refers to the node spacing and the second to the partial suppression of the current at the minimums. The node spacing in the samples takes place at $\Delta H$ ~30 G. Using $\phi_0 = \Delta H \cdot A_{effective} = 30 \cdot L \cdot d$ with $L = 20$ μm, we obtain $d$ ~35 nm. This value is in agreement with the total thickness of the junctions considering electrodes with thicknesses of ~16 nm and an STO of ~1-2 nm. The result indicates that in the limit of penetration depth $\lambda \gg d$, the value of $\Delta H$ could be tuned by changing the thickness of the electrodes. It is important to note that for bulk and optimal doped samples, $\lambda^{GBCO}(0)$ ~120 nm. This value increases as $T_s$ is reduced. Following we will discuss the partial suppression of the current at the nodes. Equation 1 predicts that the critical current vanishes for $\phi = n \cdot \phi_0$. However, a residual current was measured at the nodes for [G-1-G] and [G-2-G] samples. The main reason for this behavior could be related to the presence of pinholes. As we have shown previously for GBCO/STO bilayers, the electrical conductivity at the border of topological defects is significantly higher than the one in the terraces[19,20]. If the borders of the 3D defects are pinholes (short circuits), $I_c$ is given by two different contributions: Josephson $I_J$ (terraces) and pinholes $I_p$ (borders of 3D defects). According to this scenario, $I_J$ decreases faster than $I_p$ when the STO thickness is increased. This agrees with the fact that our measurements indicate that $I_c \sim I_p$ for [G-3-G] (see below).

It is important to mention that the IV curves for [G-3-G] display a different behavior than [G-1-G] and [G-2-G] (see Fig. 3). No Josephson coupling was observed for this sample, suggesting that the measured superconducting current originates in the electrodes and propagates through pinholes. When the polarization is up, the curves show a step-like transition to a normal state (current is injected from the top to the bottom electrode), however, when the polarization is inverted the normal state remains until a characteristic voltage where a reentrant superconducting state is observed. Suppression of the superconducting state with polarization has been previously observed in superconducting/ferroelectric (FE) bilayers[23]. It is known that ferroelectricity can appear in strained STO[24]. In this scenario, the jumps for the increment / decrement of the resistance could be originated by the switching of the FE domains[25]. The asymmetric properties for the polarization could be related to different $T_s$ for the bottom and top electrodes. Although in [G-3-G] the IV curve is dominated by pinholes ($I_p \sim I_c$), the results suggest that in JJ with FE barriers the $I_J$ could be tuned by changing the polarization. It is important to note, that no features related to FE are observed in [G-4-G], which may be related to the fact that usually the strain at the interfaces relaxes at around 3 nm[26].

Figure 4 shows the temperature dependence of $I_c R_n$ for [G-1-G] and [G-2-G]. For SIS with identical superconductors, the theoretical temperature dependence of $I_c R_n$ is limited by the gap as

$$I_c R_n = \frac{\pi \Delta(T)}{2e} tanh \frac{\Delta(T)}{2kT}, \text{[eq. 2]}$$

where $\Delta(T) = \Delta(0) tanh \left\{ 1.82 \left[ 1.018 \left( \frac{T_c}{T} - 1 \right) \right]^{0.51} \right\}$ [27] and $\Delta(0) = 1.76 k T_c$. It is important to note that for $d$-wave superconductors the $I_c(T)$ curves could display a non-monotonic behavior due to changes in the superconducting lobes that contribute to the tunneling[28] (not considered in our analysis). The results show that at 12 K (our lower temperature), the $I_c R_n$ value is strongly affected by the barrier thickness and by the reduction in the $T_s$ of the electrodes. The [G-1-G] displays $I_c R_n$ (12 K) = 4.2 mV and systematically decreases to disappear close to 77 K. [G-2-G]

displays $I_cR_n$ (12 K) = 8.9 mV and systematically decreases to disappear at approximately 44 K. A possible explanation for the reduction in its value for [G-1-G] could be related with the barrier uniformity covering the base electrode. A not uniform covering of the electrode may result in junction areas with thick barrier that does not contribute to the total $I_J$ but provides a resistive shunt to the junction. Similar behavior ($I_cR_n$ lower than the theoretical value) has been observed in $MgB_2/TiB_2/MgB_2$ junctions[10]. In our case, the performance of the junction is significantly improved for 2 nm STO barriers where a higher covering of the electrode is expected.

The value of $I_cR_n$ (12 K) = 8.9 mV observed for [G-2-G] is close to de predictions for SIS for superconductors with $T_s$ ~40 K. The value is much larger than that generally observed in $NbN$[29] and $MgB_2$[30] tunnel junctions. Moreover, the obtained value is larger than the ones usually observed in planar TJ fabricated using YBCO[31,32]. The results suggest that larger values of $I_cR_n$ could be achieved increasing the $T_c$ of the bottom electrode. The large Josephson energy is evidenced from the IV curves. High $I_cR_n$ values are of great importance to increase the working frequency for RSFQ devices and superconducting detectors. Using High-$T_c$ with high Josephson energies allows reaching the Terahertz regime with time-resolved sub picoseconds detection. Additionally, increasing the Josephson energy is fundamental to overcome temperature limitations in the operation of superconducting devices.

In summary, the presence of Josephson coupling is reported for GBCO/STO/GBCO tunnel junctions for STO barriers with thicknesses of 1 nm and 2 nm. The $I_cR_n$ values are affected by the presence of pinholes and fluctuations in the barrier thickness. We found that the $I_c$ could be divided in the contribution of Josephson coupling and pinholes. For thin insulator barriers, the presence of short circuits tends to reduce the $I_cR_n$ value. The $I_cR_n$ (12 K) = 8.9 mV observed for 2 nm thick STO is close to the theoretical prediction for TJ with $T_c$ ~40 K. Our results are promising for the development of JJ based electrical circuits going to the terahertz regime and reducing the influence of thermal noise.

FIGURES.

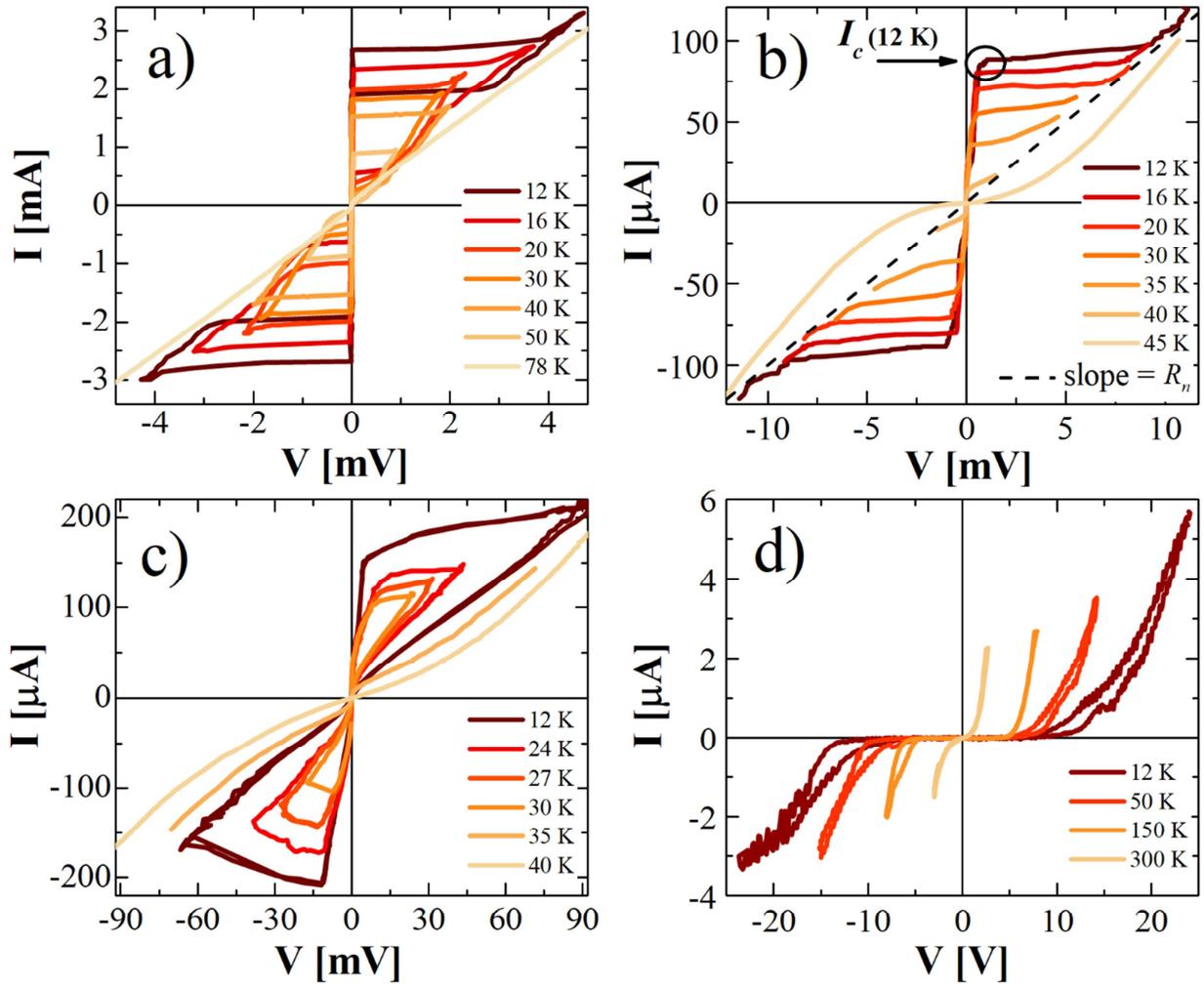

**Figure 1.** Characteristic I-V curves at different temperatures for a) [G-1-G]; b) [G-2-G]; c) [G-3-G]; and d) [G-4-G]. The curves are obtained applying current and measuring voltage. Josephson coupling is observed for JJ with STO barriers of 1 nm and 2 nm. The criteria for the determination of $I_c$ and $R_n$ are indicated in b).

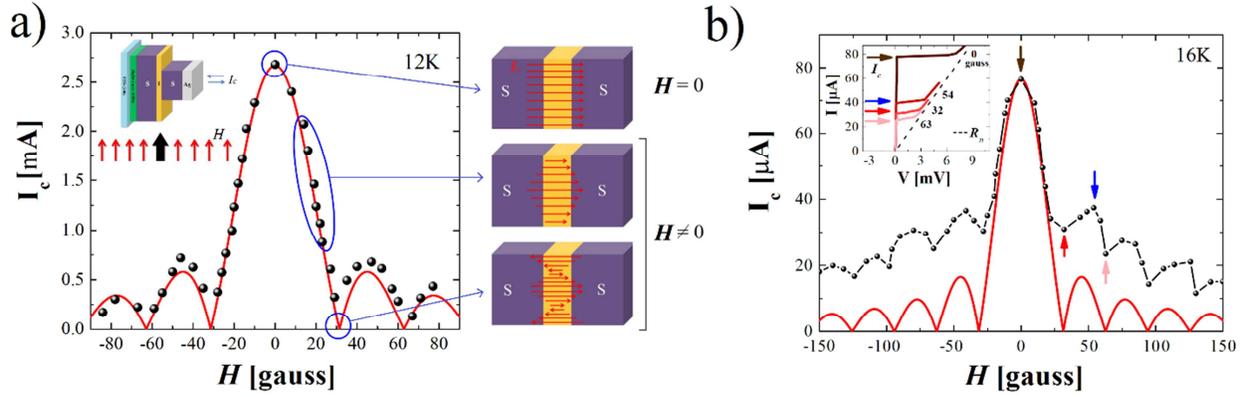

**Figure 2.** $I_c$ modulation by an external magnetic field for: a) [G-1-G]; b) [G-2-G]. The straight lines correspond to the fit for an ideal Josephson junction using equation 1. Inset a) shows a schematic tunnel junction and the magnetic field configuration. The right part of the panel a) shows a schematic representation of the effect of the magnetic field on the tunneling currents in a uniform junction. Inset b) shows typical I-V curves obtained at 16 K for different applied magnetic fields. The curves are obtained fixing the magnetic field, applying a current and measuring the voltage.

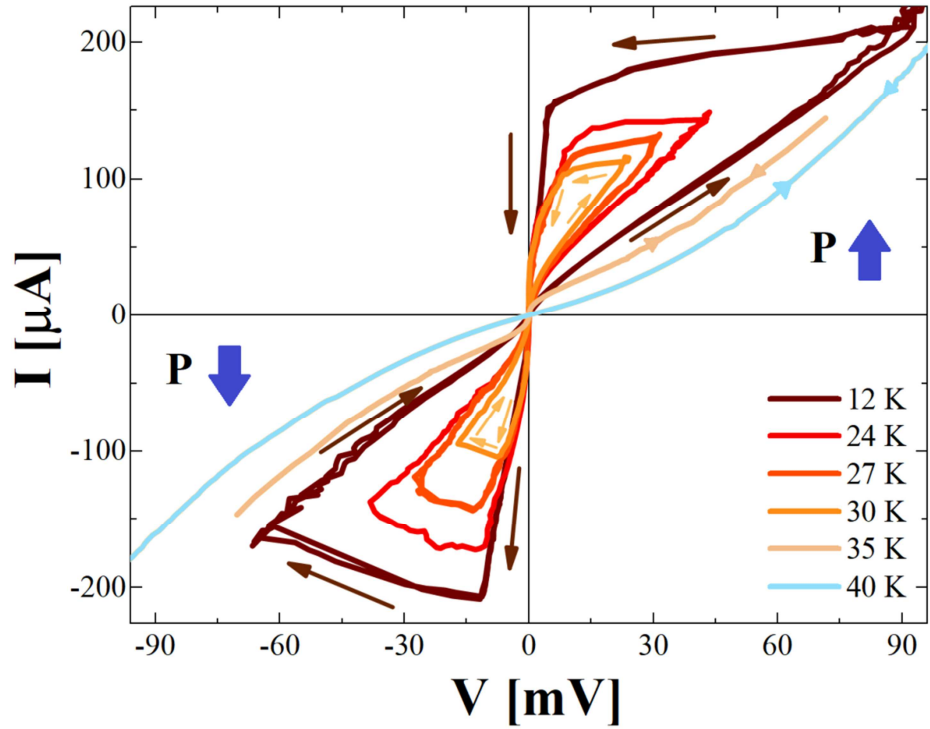

**Figure 3.** Current versus voltage measurements for [G-3-G] at different temperatures. The arrows indicate the path in which the current is applied. P indicates the direction of the polarization (up: from bottom electrode to top electrode; down: from top electrode to bottom electrode). The curves are obtained applying current and measuring voltage. Hysteretic behavior is observed for temperatures below 30 K.

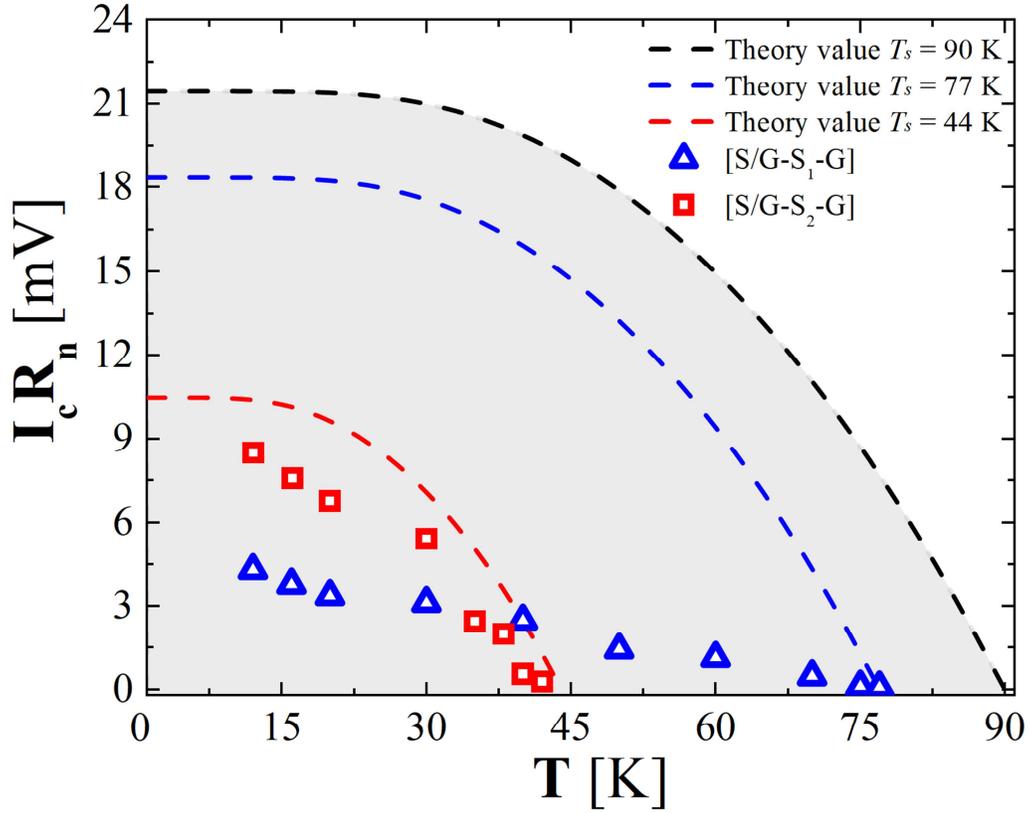

**Figure 4.** Comparison of $I_cR_n$ measured in [G-1-G] and [G-2-G] with the theoretical expectations according to the equation 2. The dashed lines correspond to the expected values considering $T_S$ of 44 K, 77 K and 90 K.

AUTHOR INFORMATION

**Corresponding Author**

*E-mail: hnavarro@physics.ucsd.edu

**Present Addresses**

†Department of Physics and Center for Advanced Nanoscience, University of California, San Diego, La Jolla, California 92093, USA.

**Author Contributions**

The manuscript was written through contributions of all authors. All authors have given approval to the final version of the manuscript.

**Funding Sources**

ANPCYT (PICT 2015-2171)

**Notes**

The authors declare no competing financial interest.

ACKNOWLEDGMENT

This work was partially supported by the ANPCYT (PICT 2015-2171). HN, MS and NH are members of the Instituto de Nanociencia y Nanotecnología CNEA-CONICET (Argentina).